\documentclass[preprint2]{aastex}

\shorttitle{progenitors for SNe Ia} \shortauthors{Meng \& Yang}

\begin{document}


\title{A comprehensive progenitor model for SNe Ia}


\author{X. Meng and W. Yang}
\affil{School of Physics and Chemistry, Henan Polytechnic
University, Jiaozuo, 454000, China} \email{xiangcunmeng@hotmail.com}






\begin{abstract}
Although the nature of the progenitor of Type Ia supernovae (SNe
Ia) is still unclear, the single-degenerate (SD) channel for the
progenitor is currently accepted, in which a carbon-oxygen white
dwarf (CO WD) accretes hydrogen-rich material from its companion,
increases its mass to the Chandrasekhar mass limit, and then
explodes as a SN Ia. The companion may be a main sequence or a
slightly evolved star (WD + MS), or a red-giant star (WD + RG).
Incorporating the effect of mass-stripping and accretion-disk
instability on the evolution of WD binary, we carried out binary
stellar evolution calculations for more than 1600 close WD
binaries. As a result, the initial parameter spaces for SNe Ia are
presented in an orbital period-secondary mass ($\log P_{\rm i},
M_{\rm 2}^{\rm i}$) plane. We confirmed that in a WD + MS system,
the initial companion leading to SNe Ia may have mass from 1
$M_{\odot}$ to 5 $M_{\odot}$. The initial WD mass for SNe Ia from WD + MS channel is
as low as $0.565$ $M_{\odot}$, while the lowest WD mass from the WD +
RG channel is 1.0 $M_{\odot}$. Adopting the results above, we
studied the birth rate of SNe Ia via a binary population synthesis
approach. We found that the Galactic SNe Ia birth rate from SD
model is $2.55 - 2.9\times10^{\rm -3}$ ${\rm yr}^{\rm -1}$ (including
WD + He star channel), which is slightly smaller than that from
observation. If a single starburst is assumed, the distribution of
the delay time of SNe Ia from the SD model may be a weak bimodality,
where WD + He channel contributes to SNe Ia with delay time shorter
than $10^{\rm 8}$ yr and WD + RG channel to those with age longer
than 6 Gyr.

\end{abstract}


\keywords{binaries: close - stars: evolution - stars: dwarf novae - supernovae: general - white dwarfs}



\section{INTRODUCTION}
\label{sect:1}
\subsection{Progenitor Models}\label{sect:1.1}
Type Ia supernovae (SNe Ia) play an important role in
astrophysics, especially in cosmology. They appear to be good
cosmological distance indicators and are successfully applied in
determining cosmological parameters (e.g. $\Omega$ and $\Lambda$),
which leads to the discovery of the accelerating expansion of the
universe (\citealt{RIE98}; \citealt{PER99}). At present, SNe Ia
are proposed to be cosmological probes for testing the evolution
of the Dark Energy equation of state with time
(\citealt{HOWEL09}). However, the progenitor systems of SNe Ia
have not yet been confidently identified (\citealt{HN00};
\citealt{LEI00}), although they show their importance in many
astrophysical field, e.g. determining cosmological parameters,
studying galaxy evolution, understanding explosion mechanism of
SNe Ia and constraining the theory of binary stellar evolution
(see the review by \citealt{LIVIO99}).

There is a consensus that SNe Ia result from the explosion of a
carbon-oxygen white dwarf in a binary system (\citealt{HF60}). The
CO WD accretes material from its companion, increases mass to its
maximum stable mass, and then explodes as a thermonuclear runaway.
Almost half of the WD mass is converted into radioactive nickel-56
in the explosion (\citealt{BRA04}), and the amount of nickel-56
determines the maximum luminosity of SNe Ia (\citealt{ARN82}).
According to the nature of the companions of the mass accreting
white dwarfs, two basic scenarios for the progenitor of SN Ia have
been discussed over the last three decades. One is the single
degenerate (SD) model (\citealt{WI73}; \citealt{NTY84}), in which
a CO WD increases its mass by accreting hydrogen- or helium-rich
matter from its companion, and explodes when its mass approaches
the Chandrasekhar mass limit. The companion may be a main-sequence or a slightly evolved
star (WD+MS) or a red-giant star (WD+RG) or a helium  star (WD + He star)
(\citealt{YUN95}; \citealt{LI97}; \citealt{HAC99a,HAC99b};
\citealt{NOM99, NOM03}; \citealt{LAN00}; \citealt{HAN04, HAN06};
\citealt{CHENWC07, CHENWC09}; \citealt{HAN08}; \citealt{MENG09};
\citealt{MENGYANG09}; \citealt{LGL09}; \citealt{WANGB09a,
WANGB09b}). An alternative is the double degenerate (DD) model
(\citealt{IT84}; \citealt{WEB84}), in which a system consisting of
two CO WDs loses orbital angular momentum by gravitational wave
radiation and merges finally. The merger may explode if the total
mass of the system exceeds the Chandrasekhar mass limit (see the
reviews by \citealt{HN00} and \citealt{LEI00}). Although the
channel is theoretically less favored, e.g. double WD mergers may
lead to accretion-induced collapses rather than to SNe Ia
(\citealt{HN00}), it is premature to exclude the channel at
present since there exists evidence that the channel may
contribute to a few SNe Ia (\citealt{HOW06}; \citealt{BRA06};
\citealt{QUI07}; \citealt{HICKEN07}; \citealt{YUAN07}). At
present, the single-degenerate Chandrasekhar model is widely
accepted, since it is supported by many observations
(\citealt{PAR07})

\subsection{Observations}\label{sect:1.2}
The SD model is supported by many observations. For example,
variable circumstellar absorption lines were observed in the
spectra of SN Ia 2006X (\citealt{PAT07}), which indicates the SD
nature of its precursor. \citet{PAT07} suggested that the
progenitor of SN 2006X is a WD + RG system based on the expansion
velocity of the circumstella material, while \citet{HKN08} argued
a WD + MS nature for this SN Ia. Recently, two twins of SN 2006X
were also reported (\citealt{BLONDIN09}; \citealt{SIMON09}), and
the birth rate and age of the 2006X-like supernovae may be well
explained by the WD + MS model (\citealt{MYG09a}). \citet{VOSS08}
suggested that SN 2007on is also possibly from a WD + MS channel.
Moreover, several SD systems are possible progenitors of SN
Ia. For example, supersoft X-ray sources (SSSs) were suggested as
good candidates for the progenitors of SNe Ia (\citealt{LIVIO96}; \citealt{KV97}; \citealt{HK03a,
HK03b}). Some of the SSSs are WD + MS systems and some are WD + RG
systems (\citealt{DIK03}). A direct way to confirm the progenitor
model is to search for the companion stars of SNe Ia in their
remnants. The discovery of the potential companion of Tycho's
supernova may have verified the reliability of the WD + MS model
(\citealt{RUI04}; \citealt{BRA04}; \citealt{IHA07}; \citealt{HERNANDEZ09}).

Measuring the delay time of SNe Ia (DD, between the episode of
star formation producing progenitor systems and the occurrence of
SNe Ia) is a very important way to constrain the progenitor
system. Although the DD of most of SNe Ia is between 0.3-2 Gyr
(\citealt{SCHAWINSKI09}), there are still SNe Ia with long delay
times (older then 10 Gyr inferred from SNe Ia in elliptical
galaxies in the local universe, \citealt{MAN05}) or extremely
short delay times (shorter than 0.1-0.3 Gyr, \citealt{MAN06};
\citealt{SCHAWINSKI09}; \citealt{RASKIN09}). Some observational
results, i.e. the strong enhancement of the SN Ia birthrate in
radio-loud early-type galaxies, the strong dependence of the SN Ia
birthrate on the colors of the host galaxies, and the evolution of
the SN Ia birthrate with redshift (\citealt{DEL05};
\citealt{MAN05, MAN06}), may indicate a bimodal distribution of delay time (DDT),
in which a part of the SNe Ia explode soon after
starburst with a delay time less than 0.1-0.3 Gyr (`prompt' SNe
Ia), while the rests have a much wider distribution with a
delay time of about 3 Gyr (`tardy' SNe Ia \citealt{MAN06}).
10\% (weak bimodality) to 50\% (strong bimodality) of all
SNe Ia belong to the prompt SNe Ia (\citealt{MANNUCCI08}).
Whether this is due to the presence of two different channels of
explosion (such as single- and double-degenerate) or to a bimodal
distribution of some parameter of the exploding systems (such as
the mass ratio between the two stars of the binary system) is
still unclear.

\subsection{Situations}\label{sect:1.3}
Many works have concentrated on the SD model. To explain prompt
SNe Ia, various models were designed. \citet{HKN08} introduced a
mass-stripping effect on a main-sequence (MS) or slightly evolved
companion star by winds from a mass-accreting white dwarf
(\citealt{HAC96}). \citet{WANGB09a, WANGB09b} investigated WD + He
star channel which can explain SNe Ia with short delay times very
well. \citet{LGL09} even designed a symbiotic channel by assuming
an aspherical stellar wind with an equatorial disk. For the tardy
SNe Ia, most of works focus on the WD + MS and WD + RG channel.
Some authors (\citealt{HAC99a, HAC99b, HKN08}; \citealt{NOM99,
NOM03}) have studied the SD channel by a simple analytical method
to treat binary interactions. Such analytic prescriptions may not
describe some mass-transfer phases, especially those occurring on
a thermal time-scale (\citealt{LAN00}). \citet{LI97} studied the
SD from detailed binary evolution calculation, while they
considered two WD masses (1.0 and 1.2 $M_{\odot}$). \citet{LAN00}
investigated the channel for metallicities $Z=0.001$ and 0.02, but
they only studied the case A evolution (mass transfer during core
hydrogen burning phase). \citet{HAN04} carried out a detailed
study of this channel including case A and early case B (mass
transfer occurs at Hertzsprung gap, HG) for $Z=0.02$. Following
the study of \citet{HAN04}, \citet{MENG09} studied the WD + MS
channel comprehensively and systematically at various Z. Based on
the works above, WD + MS channel is a very important channel to
produce SNe Ia, but the channel may only account for about 1/3 SNe
Ia observed and the delay time for the channel is between 0.2-2
Gyr (\citealt{HAN04}; \citealt{MENG09}). Recently, \citet{XL09}
proposed a WD + MS model with a thermal instable accretion disk,
which can also produce SNe Ia with long delay time. Although WD +
RG channel has been widely studied, it should be investigated
carefully by detailed binary evolution calculations with the
latest input physics since either the studies are based on a
simple analytical method or focus on some special WD masses. In
this paper, we want to construct a complete model, which includes
WD + MS and WD + RG channels.

In Section \ref{sect:2}, we describe binary evolution model, and
present the calculation results in Section \ref{sect:3}. We describe our binary
population synthesis (BPS) method in Section \ref{sect:4} and show the BPS
results in Section \ref{sect:5}. We give discussions in Section \ref{sect:6}
and summarize our main conclusions in Section \ref{sect:7}.

\begin{figure*}
\centerline{\includegraphics[angle=270,scale=.7]{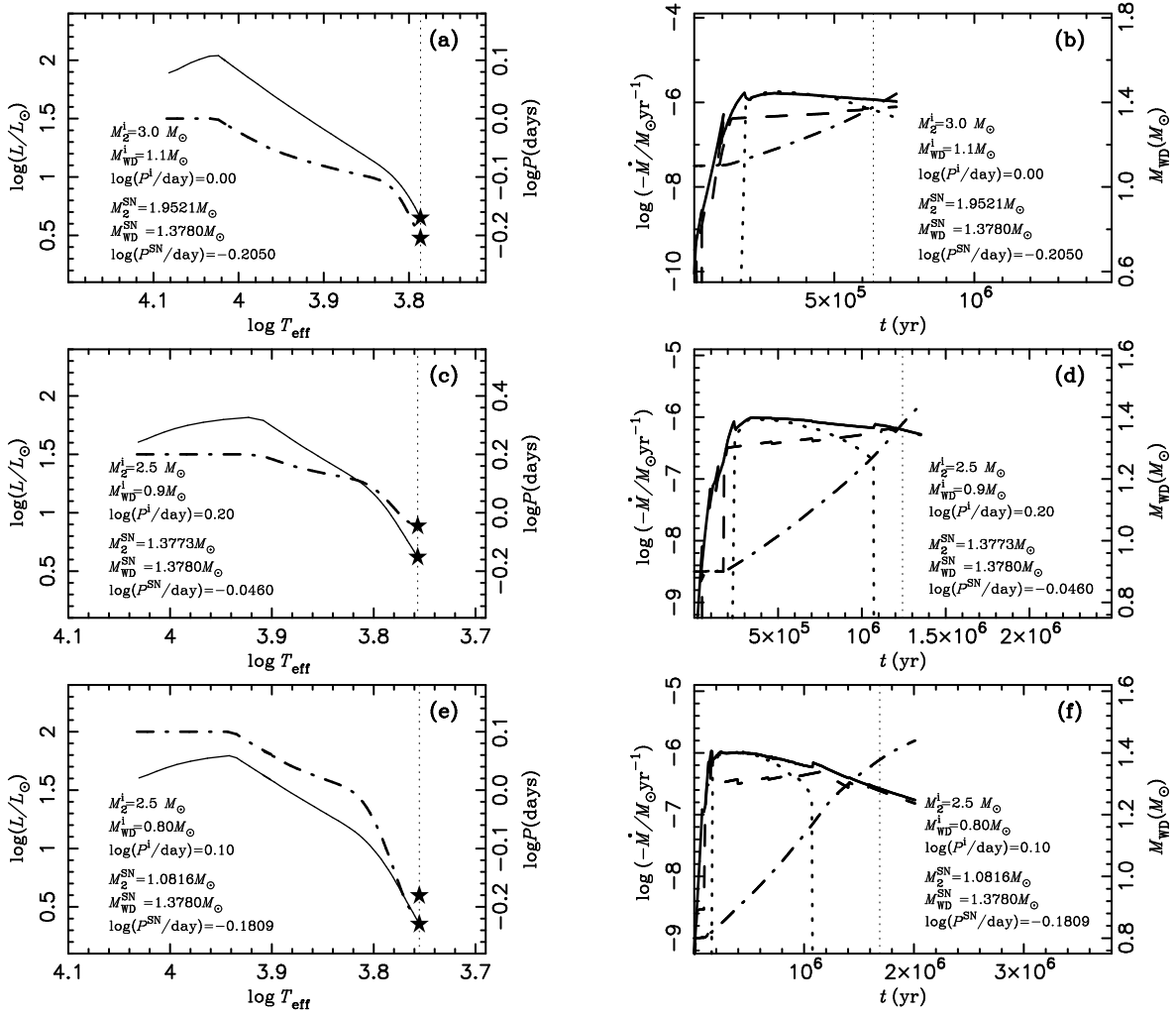}}
\caption{Three representative examples of binary evolution
calculations. The solid, dashed, dash-dotted and dotted curves
show the mass-transfer rate, $\dot{M}_{\rm tr}$, the mass-growth
rate of the CO WD, $\dot{M}_{\rm WD}$, the mass of the CO WD,
$M_{\rm WD}$, and stripping-off mass-lose rate from secondary,
$\dot{M}_{\rm strip}$, respectively, in panels (b), (d) and (f).
Dotted vertical lines in all panels and asterisks in panels (a),
(c) and (e) represent the position where the WD is expected to
explode as a SN Ia.}\label{lummdot1}
\end{figure*}

\section{BINARY EVOLUTION CALCULATION}
\label{sect:2}

\subsection{Physical Input}\label{sect:2.1}
 We use the stellar evolution code of
\citet{EGG71, EGG72, EGG73} to calculate the binary evolutions of
SD systems. The code has been updated with the latest input
physics over the last three decades (\citealt{HAN94};
\citealt{POL95, POL98}). Roche lobe overflow (RLOF) is treated
within the code described by \citet{HAN00}. We set the ratio of
mixing length to local pressure scale height, $\alpha=l/H_{\rm
p}$, to 2.0, and set the convective overshooting parameter,
$\delta_{\rm OV}$, to 0.12 (\citealt{POL97}; \citealt{SCH97}),
which roughly corresponds to an overshooting length of $0.25
H_{\rm P}$. The solar metallicity is adopted here ($Z=0.02$).
The opacity table for the metallicty is compiled by
\citet{CHE07} from \citet{IR96} and \cite{AF94}.

\subsection{Accretion Disk \& Accretion Rates}\label{sect:2.2}
In an initial binary system, the companion fills its Roche lobe at
MS or during HG or RG, and mass transfer occurs. If the mass
transfer is dynamically stable, the transferred material forms a
disk surrounding the WD. If the effective temperature in the disk
is higher than the hydrogen ionization temperature ($\sim6500$ K),
the disk is thermally stable, otherwise an unstable disk is
expected (\citealt{OSAKI96}; \citealt{VANP96}). This
corresponds to a critical mass-transfer rate, below which the disk
becomes unstable. The critical mass-transfer rate depends on the
orbital period of binary system:
 \begin{equation}
 \dot{M}_{\rm c, th}=4.3\times 10^{\rm -9}(\frac{P_{\rm orb}}{4{\rm hr}})^{1.7} M_{\odot}{\rm yr}^{\rm -1},
  \end{equation}
where $P_{\rm orb}$ is the orbital period unit in hour
(\citealt{OSAKI96}; \citealt{VANP96}). When mass-transfer rate
$|\dot{M}_{\rm tr}|$ exceeds the critical value, $\dot{M}_{\rm c,
th}$, we assume the accretion disk is stable and the WD
accretes the transferred material smoothly at a rate $\dot{M}_{\rm
a}=|\dot{M}_{\rm tr}|$. If $|\dot{M}_{\rm tr}|$ is lower than
$\dot{M}_{\rm c, th}$, mass is stored within the disk and no
material is accreted by WD. When the materials stored in the disk
reaches some critical amount, they are suddenly accreted onto the
WD at a rate $\dot{M}_{\rm a}=|\dot{M}_{\rm tr}|/d$ due to thermal
instability , which may explain dwarf-nova outburst excellently
(see the review by \citealt{OSAKI96} for details about the
disk-instability model). Here, $d$ is duty cycle and is set to be
0.01 as did by \citet{XL09}. Then, the accretion rate of WD is
  \begin{equation}
\dot{M}_{\rm a}=\left\{
 \begin{array}{ll}
 |\dot{M}_{\rm tr}|, & |\dot{M}_{\rm tr}|\geq\dot{M}_{\rm c, th},\\
 |\dot{M}_{\rm tr}|/d, & |\dot{M}_{\rm tr}|<\dot{M}_{\rm c, th},\\
\end{array}\right.
\end{equation}
where the typical timescale for the second accretion rate
(in the case of the duty cycle) is equal to $M_{\rm disk}/\dot{M}_{\rm a}$, where
$M_{\rm disk}$ is the mass accumulated in the disk (see review by \citet{LASOTA01}
in details).

\subsection{WD Mass Growth}\label{sect:2.3}
Instead of solving stellar structure equations of a WD, we adopt
the prescription of \citet{HAC99a} and \citet{HKN08} on WDs
accreting hydrogen-rich material from their companions. The
following is a brief introduction of this prescription. If WD
accretion rate, $\dot{M}_{\rm a}$, exceeds another critical value,
$\dot{M}_{\rm c, H}$, we assume that a part of accreted hydrogen
steadily burns on the surface of WD and
is converted into helium at the rate of $\dot{M}_{\rm c,
H}$. The unprocessed matter is assumed to be lost from the system
as an optically thick wind at a rate of $\dot{M}_{\rm
wind}=\dot{M}_{\rm a}-\dot{M}_{\rm c, H}$ (\citealt{HAC96}). The
optically thick wind may compact into the envelope of WD companion
and strips off some hydrogen-rich material from the companion
surface. The mass-stripping effect may attenuates the
mass-transfer rate from the companion to the WD, and was
successfully adapted to explain some quasi-regular SSSs (\citealt{HK03a,HK03b}). The
mass-lose rate for the mass-stripping effect, $\dot{M}_{\rm
strip}$ is proportional to $\dot{M}_{\rm wind}$:
 \begin{equation}
 \dot{M}_{\rm strip}=c_{\rm 1}\dot{M}_{\rm wind},
  \end{equation}
where $c_{\rm 1}$ a constant. At present, the value of $c_{\rm 1}$
is very uncertain. To explain the position of a SSS (V Sge) in
the initial orbital period-secondary mass ($\log P^{\rm i}, M_{\rm
2}^{\rm i}$) plane, $c_{\rm 1}$ should be larger than 0
(\citealt{HKN08}). \citet{HKN08} checked the influence of $c_{\rm
1}$ on the donor mass range and found that the range increases
with $c_{\rm 1}$. Considering primordial primary is more massive
than primordial secondary, the $c_{\rm 1}$ should be smaller 3
(see Fig. 3 in \citealt{HKN08} and the discussion in Section \ref{subs:5.1}).
In this paper, we assume rather
arbitrarily that $c_{\rm 1}=1.5$, which is the intermediate value
between 0 and 3. Then, the mass-lose rate of the companion is
$\dot{M}_{\rm 2}=\dot{M}_{\rm tr}-\dot{M}_{\rm strip}$\footnote{Please note that $\dot{M}_{\rm 2}$
and $\dot{M}_{\rm tr}$ are negative, while $\dot{M}_{\rm strip}$ is positive.}. The
material lost from the system may form circumstellar material (CSM),
which is a possible origin of color excess of SNe Ia
(\citealt{MENGXC09}).

The critical accretion rate is
 \begin{equation}
 \dot{M}_{\rm c, H}=5.3\times 10^{\rm -7}\frac{(1.7-X)}{X}(M_{\rm
 WD}-0.4) M_{\odot} {\rm yr}^{\rm -1},
  \end{equation}
where $X$ is hydrogen mass fraction and $M_{\rm WD}$ is the mass
of the accreting WD (mass is in $M_{\odot}$ and mass-accretion
rate is in $M_{\odot}/{\rm yr}$, \citealt{HAC99a}).

The following assumptions are adopted when $\dot{M}_{\rm a}$ is
smaller than $\dot{M}_{\rm c, H}$. (1) When $\dot{M}_{\rm a}$ is
higher than $\frac{1}{2}\dot{M}_{\rm c, H}$, the hydrogen-shell
burning is steady and no mass is lost from the system. (2) When
$\dot{M}_{\rm a}$ is lower than $\frac{1}{2}\dot{M}_{\rm c, H}$
but higher than $\frac{1}{8}\dot{M}_{\rm c, H}$, a very weak shell
flash is triggered but no mass is lost from the system. (3) When
$\dot{M}_{\rm a}$ is lower than $\frac{1}{8}\dot{M}_{\rm c, H}$,
the hydrogen-shell flash is so strong that no material is
accumulated on to the surface of the CO WD\footnote{Actually,
when a WD undergos nova explosions,
its mass is reduced. However,
this should not affect our conclusions about SNe Ia.}. We define the growth rate
of the mass of the helium layer under the hydrogen-burning shell
as
 \begin{equation}
 \dot{M}_{\rm He}=\eta _{\rm H}\dot{M}_{\rm a},
  \end{equation}
where $\eta _{\rm H}$ is the mass accumulation efficiency for
hydrogen burning. According to the assumptions above, the values
of $\eta _{\rm H}$ are:

 \begin{equation}
\eta _{\rm H}=\left\{
 \begin{array}{ll}
 \dot{M}_{\rm c, H}/\dot{M}_{\rm a}, & \dot{M}_{\rm a}> \dot{M}_{\rm
 c, H},\\
 1, & \dot{M}_{\rm c, H}\geq \dot{M}_{\rm a}\geq\frac{1}{8}\dot{M}_{\rm
 c, H},\\
 0, & \dot{M}_{\rm a}< \frac{1}{8}\dot{M}_{\rm c, H}.
\end{array}\right.
\end{equation}

Helium is ignited when a certain amount of helium is accumulated.
If a He-flash occurs, some of the helium is blown off from the
surface of the CO WD. Then, the mass growth rate of the CO WD,
$\dot{M}_{\rm WD}$, is
 \begin{equation}
 \dot{M}_{\rm WD}=\eta_{\rm He}\dot{M}_{\rm He}=\eta_{\rm He}\eta_{\rm
 H}\dot{M}_{\rm a},
  \end{equation}
where $\eta_{\rm He}$ is the mass accumulation efficiency for
helium-shell flashes, and its value is taken from \citet{KH04}.
\\
\\
We incorporated all the above prescription into Eggleton's stellar
evolution code and followed the evolutions of both the mass donor
and the accreting CO WD. The mass lost as optically thick wind is
assumed to take away the specific orbital angular momentum of the
accreting WD, while the stripped-off material by the optically
thick wind is assumed to take away the specific orbital angular
momentum of donor star. We calculated about 1600 binary systems,
and obtained a large, dense model grid. $M_{\rm 2}^{\rm i}$ range
from 0.6 $M_{\odot}$ to 5 $M_{\odot}$; the initial masses of CO
WDs, $M_{\rm WD}^{\rm i}$, from 0.565 $M_{\odot}$ to 1.20
$M_{\odot}$; the initial orbital periods of binary systems,
$P^{\rm i}$, from the minimum value (at which a zero-age
main-sequence (ZAMS) star fills its Roche lobe) to $\sim 80$ days.
In the calculations, we assume that the WD explodes as a SN Ia
when its mass reaches the Chandrasekhar mass limit, i.e. 1.378
$M_{\odot}$




\section{BINARY EVOLUTION RESULTS}\label{sect:3}
 \begin{figure*}
 \centerline{\includegraphics[angle=270,scale=.70]{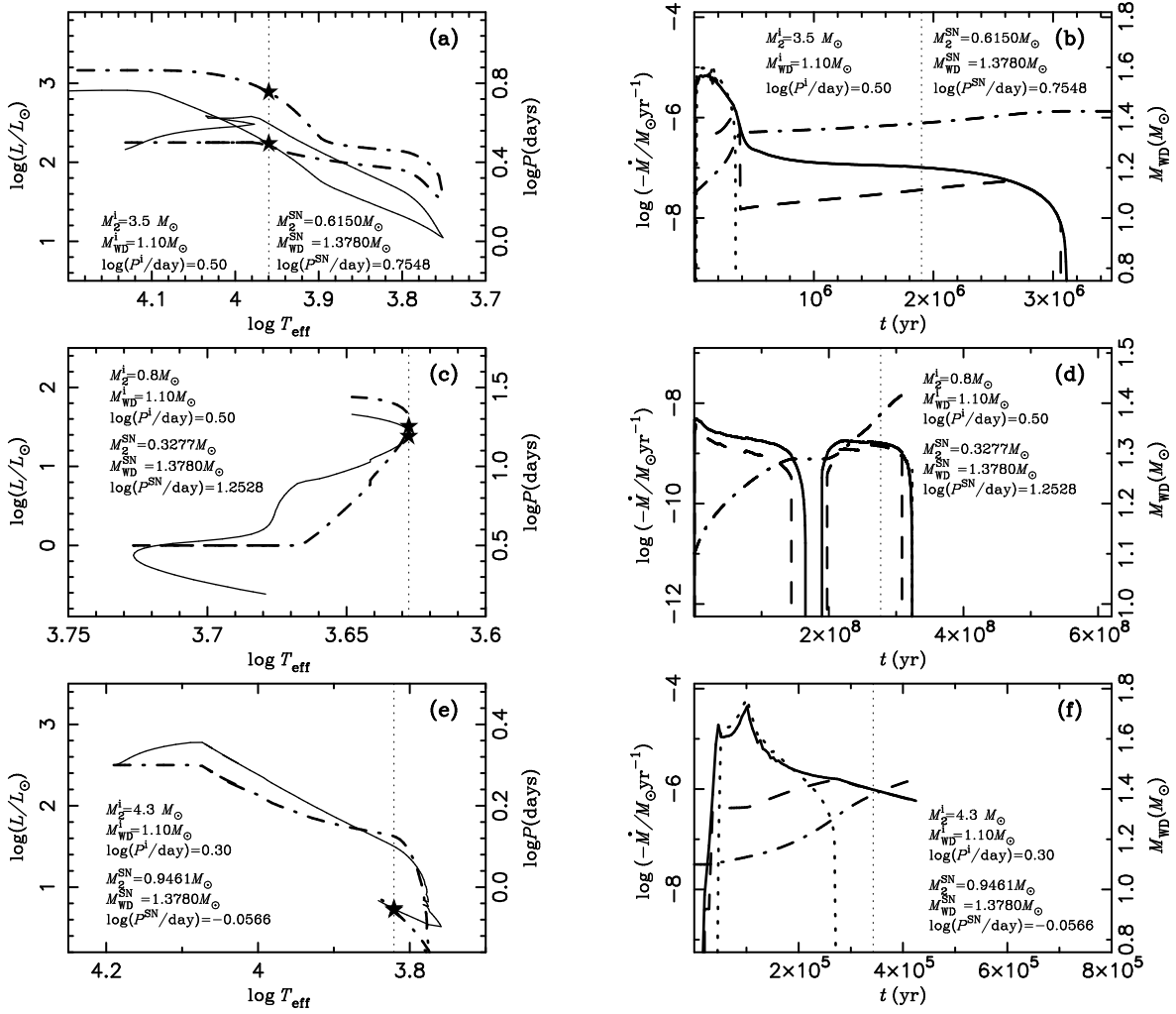}}
 \caption{Similar to Fig. \ref{lummdot1} but for two extreme
 examples and a WD + RG one.}\label{lummdot2}
 \end{figure*}

In Figs. \ref{lummdot1} and \ref{lummdot2}, we present four
representative examples and two extreme examples of
our binary evolution calculations. The figures show the mass-transfer rate,
$\dot{M}_{\rm tr}$, the growth rate of CO WD, $\dot{M}_{\rm
WD}$, the mass of CO WD, $M_{\rm WD}$, stripping-off mass-lose
rate, $\dot{M}_{\rm strip}$, the evolutionary track of donor
star in the Hertzsprung-Russell (HR) diagram and the evolution of
orbital period.

\subsection{Case Wind}\label{sect:3.1}
Panels (a) and (b) in Fig. 1 represent the evolution of a binary
system with an initial mass of the donor star of $M_{\rm 2}^{\rm
i}=3.0M_{\odot}$, an initial mass of the CO WD of $M_{\rm WD}^{\rm
i}=1.10M_{\odot}$ and an initial orbital period of $\log(P^{\rm
i}/{\rm day})=0.00$. The donor star fills its Roche lobe at the MS
stage which results in a case A RLOF. The mass-transfer rate
exceeds $\dot{M}_{\rm c, H}$ soon after the onset of the RLOF.
This results in a wind phase, where a part of the transferred
mass is blown off as an optically thick wind, while the rest is
accumulated on to the WD. The optically thick wind collides into the
envelope of the companion and strips off some hydrogen-rich
material from the surface of the companion. When the mass reaches
$M_{\rm WD}^{\rm SN}=1.378 M_{\odot}$,  where the WD is assumed to
explode as a SN Ia, the system is still in the wind phase (see the thick dotted line).
At this point, the mass of the companion is $M_{\rm 2}^{\rm SN}=1.9521
M_{\odot}$ and the orbital period is $\log(P^{\rm i}/{\rm
day})=-0.2050$. Since the mass-transfer rate from the secondary
continuously exceeds $\dot{M}_{\rm c, H}$ until the WD explodes as
an SN Ia, we call this ``Case Wind" as did in \citet{HKN08} and
\citet{MYG09b}. Case Wind is realized in the region of $2.3
M_{\odot}\leq M_{\rm 2}^{\rm i}\leq3.8M_{\odot}$ and $P^{\rm
i}\leq5$ days for $M_{\rm WD}^{\rm i}=1.0$ and 1.1 $M_{\odot}$. As
noticed by \citet{HKN08} and \citet{MYG09b}, no Case Wind exists
for $M_{\rm WD}^{\rm i}<0.9 M_{\odot}$ (see also Figs. \ref{grid2} and \ref{grid1}).

Before supernova explosion, Case Wind may be observed as quasi-regular
transient SSS such as V Sge (\citealt {KATO09}).
The hydrogen-rich material lost from the system may form CSM and
contribute to the intrinsic color excess of SNe Ia (\citealt{MENGXC09}). The material
loses from the system via two ways: as optically thick wind from the
surface of WD or stripped-off material by the wind from the companion. The
velocity of wind should be larger than the escape velocity of the
WD ( an order of magnitude of $10^{\rm 3}$ ${\rm km}$ ${\rm s^{\rm
-1}}$), which is too large to contribute the intrinsic color
excess of SNe Ia (\citealt{MENGXC09}).
The velocity of the stripped-off material is relatively
low ($\sim 100$ ${\rm km}$ ${\rm s^{\rm -1}}$, \citealt{HKN08}).
Since the WD explodes during the wind phase,
the stripped-off material may form CSM very close the
SN Ia and may be the possible origin of supernovae like
SN 2002ic and SN 2006X (\citealt{HKN08}; \citealt{MYG09a}). It
may be helpful for explaining the properties of SNe Ia remnants
(\citealt{BADENES07}).

\subsection{Case Calm}\label{sect:3.2}
Panels (c) and (d) in Fig. 1 show another example for an initial
system with $M_{\rm 2}^{\rm i}=2.50M_{\odot}$, $M_{\rm WD}^{\rm
i}=0.90M_{\odot}$, $\log(P^{\rm i}/{\rm day})=0.20$. Similar to
the previous example, the companion also fills its Roche lobe at
the MS stage and the system experiences a wind phase after the
onset of the RLOF. After mass ratio reverse, $|\dot{M}_{\rm tr}|$
drops until below $\dot{M}_{\rm c, H}$, but is still higher than
$\frac{1}{2}\dot{M}_{\rm c, H}$, i.e. the optically thick wind stops
and hydrogen-shell burning is stable. During this phase, WD reaches
$M_{\rm WD}^{\rm SN}=1.378 M_{\odot}$, where $M_{\rm 2}^{\rm
SN}=1.3773 M_{\odot}$ and $\log(P^{\rm i}/{\rm day})=-0.0460$. The
main difference between this example and Case Wind is that
at the time of the explosion, the system is in the stable
hydrogen-burning phase after the optically thick wind phase while
Case Wind is still in wind phase. Since WD undergoes steady
hydrogen burning at the time of SN Ia explosion, we call this
``Case Calm" as did in \citet{HKN08} and \citet{MYG09b}. Case Calm
can be realized when $M_{\rm WD}^{\rm i}\geq0.80M_{\odot}$
(see also Figs. \ref{grid2} and \ref{grid1}).

Before supernova explosion, Case Calm may be observed as
persistent SSS (\citealt{HKN08}; \citealt{KATO09}). The materials
lost as wind or stripped off by the wind form CSM, but they have
been dispersed too far to be detected immediately after SN Ia
explosion. We even can not expect radio or X-ray emission until at
lease 10 - 100 yr after the explosion. Then, Case calm would show
the properties of ``normal'' SNe Ia (\citealt{BRA93};
\citealt{HKN08}).

\subsection{Case Nova}\label{sect:3.3}
The third example in panels (e) and (f) of Fig. 1 represents an
initial system with $M_{\rm 2}^{\rm i}=2.50M_{\odot}$, $M_{\rm
WD}^{\rm i}=0.80M_{\odot}$ and $\log(P^{\rm i}/{\rm day})=0.10$.
Its evolution is similar to that of Case Calm except that the WD
explodes at the phase where $\frac{1}{2}\dot{M}_{\rm c,
H}\geq|\dot{M}_{\rm tr}|\geq\frac{1}{8}\dot{M}_{\rm c, H}$. When $M_{\rm
WD}^{\rm SN}=1.378 M_{\odot}$, $M_{\rm 2}^{\rm SN}=1.0816
M_{\odot}$ and $\log(P^{\rm i}/{\rm day})=-0.1809$. Before
supernova explosion, hydrogen-burning is unstable, and the system may
be observed as a recurrent nova (\citealt{HKN08};
\citealt{KATO09}). Therefore, we call this ``Case Nova" as did in
\citet{HKN08} and \citet{MYG09b}. Case Nova can be realized when
$M_{\rm WD}^{\rm i}\geq0.70M_{\odot}$\footnote{Please note that
the WD masses in Figs. \ref{grid2} and \ref{grid1} are the initial masses of WDs, while the cases of
``Case Wind", ``Case Calm" and ``Case Nova" are defined according to the evolutional stage of
a WD binary system at the moment of supernova explosion.}.

The materials lost as wind or stripped off by the wind form CSM,
but they have been dispersed too far to be detected immediately
after the SN Ia explosion. It takes at lease 100 yr for supernova
ejecta to reach the CSM. So, a ``normal'' SN Ia is expected
(\citealt{BRA93}; \citealt{HKN08}).

\subsection{Case DNova}\label{sect:3.4}
Panels (a) and (b) in Fig. 2 represent the evolution of a binary
system with an initial mass of the donor star of $M_{\rm 2}^{\rm
i}=3.5M_{\odot}$, an initial mass of the CO WD of $M_{\rm WD}^{\rm
i}=1.10M_{\odot}$ and an initial orbital period of $\log(P^{\rm
i}/{\rm day})=0.50$. After RLOF occurs, the system experiences winds phase soon. Then,
the system lost much hydrogen-rich material.
After mass-ratio reverse, the system experiences the wind, the
stable and weakly unstable hydrogen-burning phase one by one, until $|\dot{M}_{\rm
tr}|<\frac{1}{8}\dot{M}_{\rm c, H}$ and even $|\dot{M}_{\rm
tr}|<\dot{M}_{\rm c, th}$. At this phase, $M_{\rm WD}$ still dose not
reach to $1.378M_{\odot}$. Then, the  WD increases its mass
gradually by accretion from a thermal unstable disk.
After about $1.3\times10^{\rm 6}$ yr, $M_{\rm WD}^{\rm
SN}$ reaches to $1.378 M_{\odot}$, where $M_{\rm 2}^{\rm
SN}=0.6150 M_{\odot}$ and $\log(P^{\rm i}/{\rm day})=0.7548$.
During this phase, the system may be observed as a dwarf nova (\citealt{OSAKI96}). So,
we call this ``Case DNova". Case DNova is mainly realized when
$M_{\rm 2}^{\rm i}/M_{\rm WD}^{\rm i}\gtrsim3$ and the system
is crossing HG at the onset of RLOF. For Case DNova, after
$|\dot{M}_{\rm tr}|<\frac{1}{8}\dot{M}_{\rm c, H}$, $|\dot{M}_{\rm tr}|$ may be still higher than
$\dot{M}_{\rm c, th}$. During this phase, the hydrogen-burning is heavily unstable, and then, the system
should show the properties of classical nova (\citealt{KATO09}).

For the same reason to Case Calm and Case Nova, a ``normal'' SN Ia is expected for
Case DNova.

\subsection{Case Instability}\label{sect:3.5}
Panels (e) and (f) in Fig. 2 illustrate a more extreme case where
both the donor and the WD are relatively massive. The initial binary in this
case are $M_{\rm WD}^{\rm i}=1.10 M_{\odot}$, $M_{\rm 2}^{\rm
i}=4.3 M_{\odot}$ and $\log(P^{\rm i}/{\rm day})=0.30$. After
about $1\times10^{\rm 5}$ yr from the onset of RLOF, mass transfer
becomes almost dynamically unstable and hence the mass-transfer
rate increases sharply, only to drop once the mass ratio has been
reversed. When $M_{\rm WD}^{\rm i}=\rm1.378 M_{\odot}$, the binary
parameters are $M_{\rm 2}^{\rm SN}=0.9461 M_{\odot}$ and
$\log(P^{\rm i}/{\rm day})=-0.0566$. For a larger initial donor
mass, e.g. $M_{\rm 2}^{\rm i}=4.5 M_{\odot}$, our calculations show that
mass transfer is unstable, and such systems experience a delayed dynamical
instability (\citealt{HW87}; \citealt{POD02}). So, we call this case ``Case Instability"
There are mainly two differences between Case Instability and Case DNova. One is that Case Instability
occurs when mass transfer begin at MS stage while Case DNova during HG. The other is that
for Case Instability, both the donor and the WD are relatively massive, while
Case DNova has no such constraint.

\subsection{Case RGB}\label{sect:3.6}
Panels (c) and (d) in Fig. 2 show a case that RLOF begin at red
giant branch (RGB). The parameters of the initial system are
$M_{\rm WD}^{\rm i}=1.10 M_{\odot}$, $M_{\rm 2}^{\rm i}=0.8
M_{\odot}$ and $\log(P^{\rm i}/{\rm day})=0.50$. For the system,
the mass transfer rate is always lower than $\dot{M}_{\rm c, H}$.
There is even a broken phase for mass transfer.
The WD mainly accretes hydrogen-rich material from a thermal unstable disk.  After about
$2.7\times10^{\rm 8}$ yr, $M_{\rm WD}^{\rm SN}$ reaches to $1.378
M_{\odot}$, where $M_{\rm 2}^{\rm SN}=0.3277 M_{\odot}$ and
$\log(P^{\rm i}/{\rm day})=1.2528$. To distinguish with Case DNova
and due to the RGB nature of the donor, we call this ``Case RGB".
Case RGB is only realized when $M_{\rm WD}^{\rm i}\geq1.00
M_{\odot}$.  In our results, all WD + RG systems explode at the phase
of $\dot{M}_{\rm tr}<\dot{M}_{\rm c, th}$, and then they should be
observed as dwarf nova before SNe Ia.

\subsection{Initial Parameters for the Progenitor of SNe Ia}\label{sect:3.7}

 \begin{figure*}
 \centerline{\includegraphics[angle=270,scale=.75]{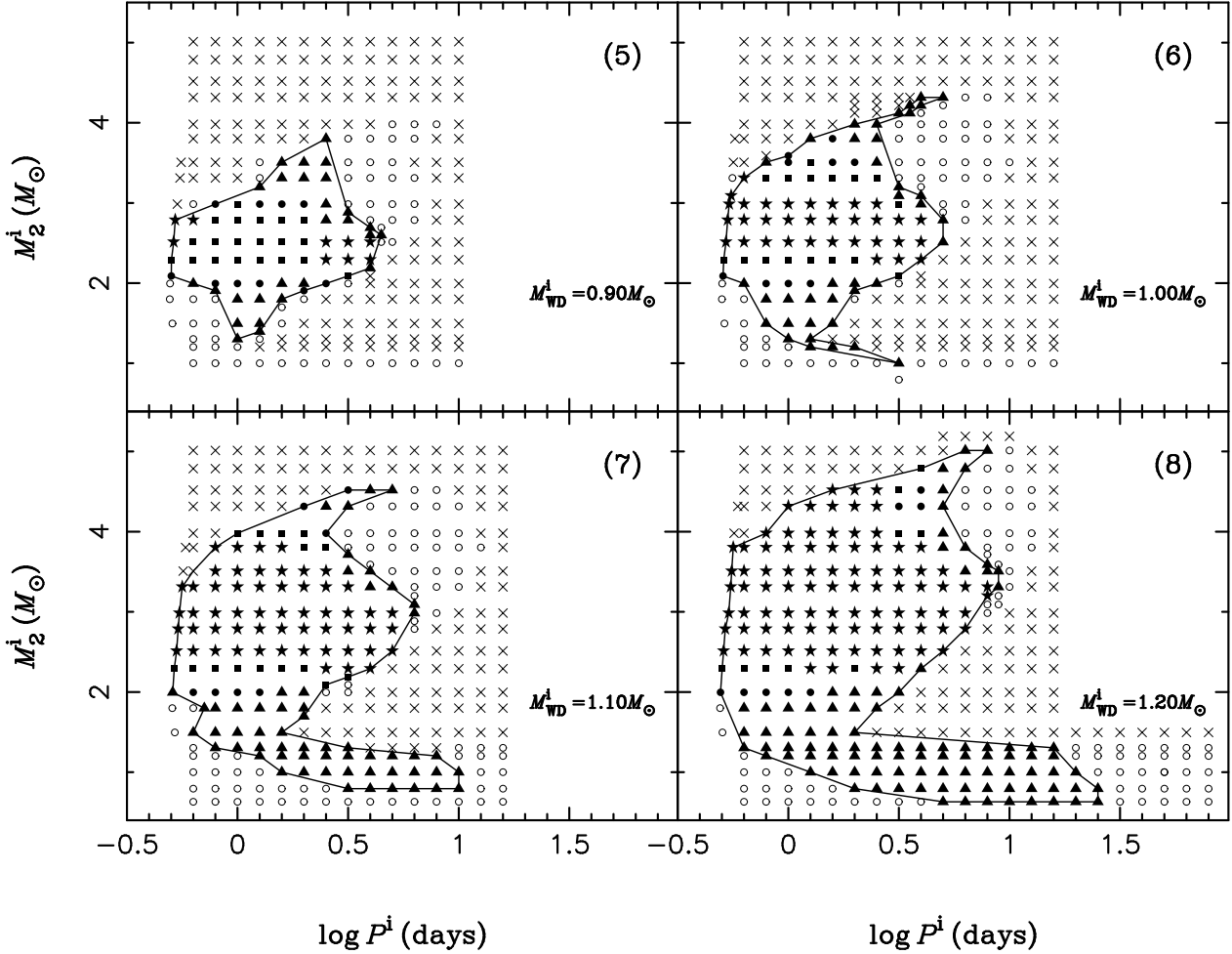}}
 \caption{Final outcomes of binary evolution calculation in the
 initial orbital period-secondary mass ($\log P^{\rm i}, M_{\rm 2}^{\rm i}$) plane,
 where $P^{\rm i}$ is the initial orbital period and $M_{\rm 2}^{\rm i}$
 is the initial mass of donor star (for different initial WD masses
 as indicated in each panel). Filled five-pointed stars indicate SN Ia
 explosions during an optically thick wind phase ($|\dot{M}_{\rm tr}|>\dot{M}_{\rm c, H}$).
 Filled squares denote SN Ia explosions after the wind phase, where hydrogen-shell burning is stable
 ($\dot{M}_{\rm c, H}\geq |\dot{M}_{\rm tr}|\geq \frac{1}{2}\dot{M}_{\rm c, H}$).
 Filled circles denote SN Ia explosions after the wind phase where hydrogen-shell burning is
 mildly unstable ($\frac{1}{2}\dot{M}_{\rm c, H}> |\dot{M}_{\rm tr}|\geq \frac{1}{8}\dot{M}_{\rm c, H}$).
 Filled triangles denote SN Ia explosions during unstable disk phase
 ($|\dot{M}_{\rm tr}|<\dot{M}_{\rm c, th}$ and $|\dot{M}_{\rm tr}|/d\geq\frac{1}{8}\dot{M}_{\rm c, H}$).
 Open circles indicate systems that experience novae
 explosion, preventing the CO WD from reaching 1.378 $M_{\odot}$ ($|\dot{M}_{\rm tr}|<\dot{M}_{\rm c, th}$ and
 $|\dot{M}_{\rm tr}|/d<\frac{1}{8}\dot{M}_{\rm c, H}$),
 while crosses show the systems that are unstable to dynamical mass transfer.
 }\label{grid2}
 \end{figure*}
\begin{figure*}
 \centerline{\includegraphics[angle=270,scale=.75]{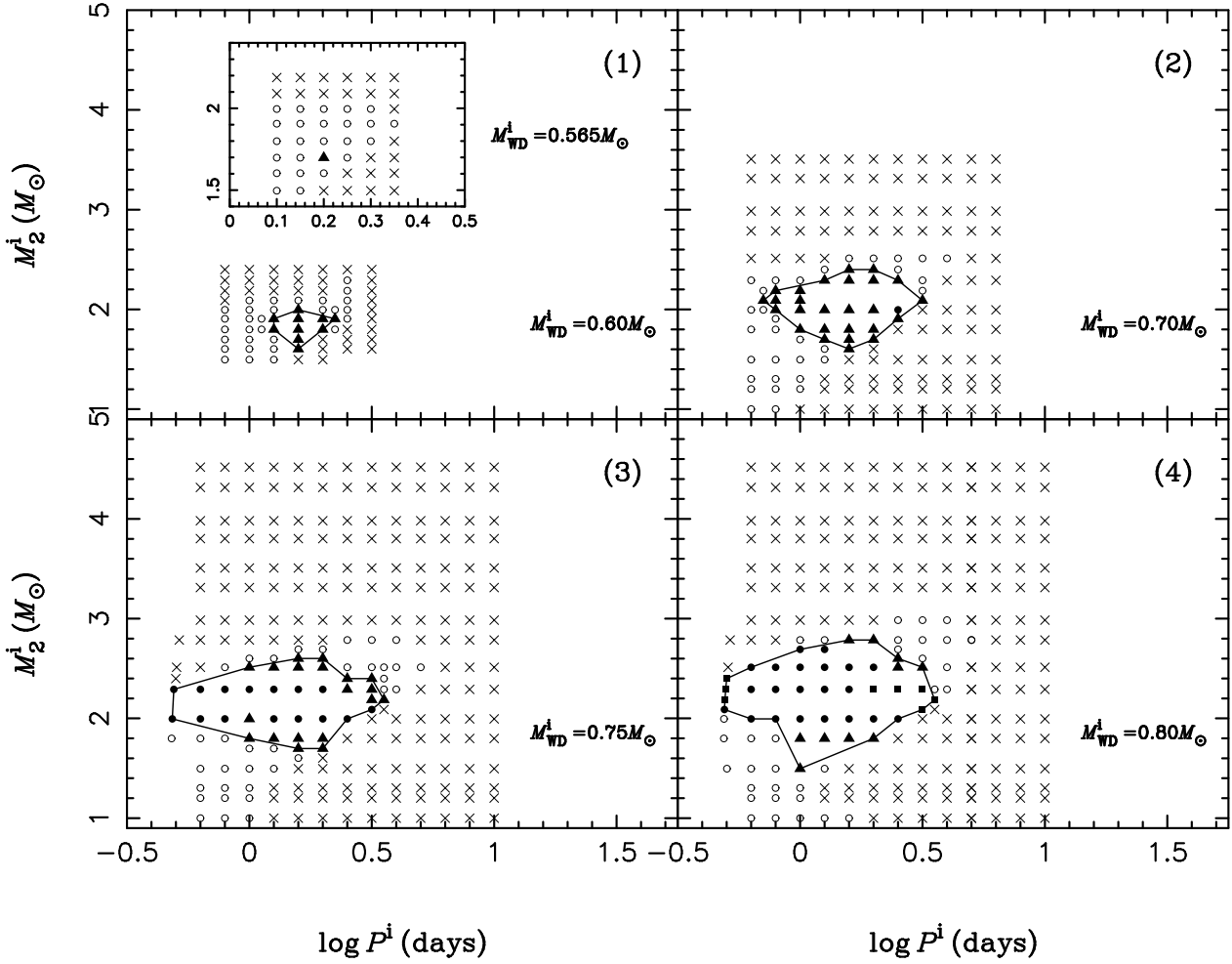}}
 \caption{Similar to Fig. \ref{grid2} but for different initial WD masses.}\label{grid1}
 \end{figure*}

To conveniently compare our results with previous studies in
literatures, we summarize the final outcomes of all the binary
evolution calculations in the initial orbital period-secondary
mass ($\log P^{\rm i}, M_{\rm 2}^{\rm i}$) plane.
Figs. \ref{grid2} and \ref{grid1} summarize the final outcome of our binary evolution
calculations in the ($\log P^{\rm i}, M_{\rm 2}^{\rm i}$) plane.
Filled symbols show the results leading to SNe Ia,
where the shape of the symbols indicates the supernova explosions
in the optically thick wind phase (filled five-pointed star: $|\dot{M}_{\rm tr}|>\dot{M}_{\rm c, H}$), after
the wind phase while in the stable hydrogen-burning phase (Filled squares:
$\dot{M}_{\rm c, H}\geq |\dot{M}_{\rm tr}|\geq \frac{1}{2}\dot{M}_{\rm c, H}$)
or in the unstable hydrogen-shell burning phase (Filled circles: $\frac{1}{2}\dot{M}_{\rm c, H}> |\dot{M}_{\rm tr}|\geq \frac{1}{8}\dot{M}_{\rm c, H}$),
and unstable disk phase (Filled triangles: $|\dot{M}_{\rm tr}|<\dot{M}_{\rm c, th}$ and $|\dot{M}_{\rm tr}|/d\geq\frac{1}{8}\dot{M}_{\rm c, H}$).
Systems experiencing
nova explosions and never reaching the Chandrasekhar limit and systems
experiencing a dynamical mass transfer are also indicated in the figures.

\begin{figure}
\centerline{\includegraphics[angle=270,scale=.35]{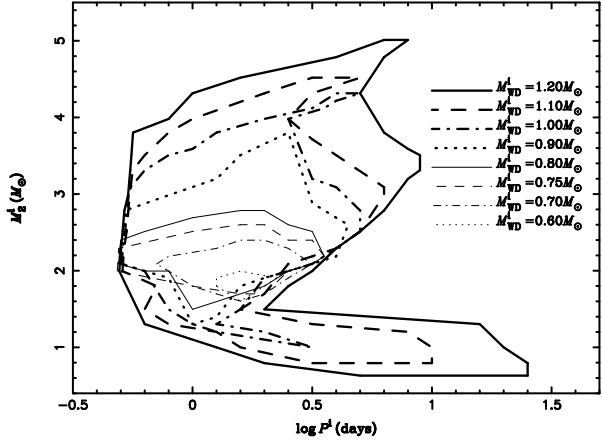}}
\caption{Regions in the initial orbital period-secondary mass
plane ($\log P^{\rm i}, M_{\rm 2}^{\rm i}$) for WD binaries
producing SNe Ia for initial WD masses of 0.60, 0.70, 0.75, 0.80,
0.90, 1.00, 1.10 and 1.20 $M_{\odot}$. Since the region almost
vanishes for $M_{\rm WD}^{\rm i}=0.565M_{\odot}$, we do not show it in the figure.}\label{contour}
\end{figure}

In Figs. \ref{grid2} - \ref{contour}, we also present the contours of the initial parameters
in which SNe Ia are expected. In the figures, the left boundaries are determined by the radii
of ZAMS stars, i.e. RLOF starts at zero age.
the WD + MS systems beyond the right boundaries will undergo dynamically
unstable mass transfer at the base of the red giant branch
(RGB), while WD + RG systems beyond the right boundaries will experience strong hydrogen-shell flash.
The upper boundaries are determined by the delayed dynamical instability or the strong hydrogen-shell flash.
The lower boundaries are constrained by the
condition that the WD mass accretion rate is larger than
$\frac{1}{8}\dot{M}_{\rm c, H}$ and that the secondaries have enough
material to transfer on to CO WDs, which can then increase their
masses to 1.378 $M_{\odot}$.

In these figures, we can  clearly see that WD + RG channel occurs
only when $M_{\rm WD}^{\rm i}\geq1.00 M_{\odot}$ and the region
for WD + RG channel increases with initial WD mass and extends to
longer period. However, the maximum initial period for WD + RG
channel is  $\sim25$ days, because the mass-transfer rate is very high and
most of transferred material loses from system as optically thick wind or
stripped-off material if the period is very large.  Similar to \citet{XL09}, the lower boundary of
WD + MS channel extends to $\sim1M_{\odot}$, because the effect of
a thermally unstable disk on the evolution of WD binaries is included in this paper.
The upper boundary for WD + MS channel move to
$\sim5M_{\odot}$, because we include the wind mass-stripping
effect suggested by \citet{HKN08}. The high donor mass results
might contribute to young population of SNe Ia.

In Fig. \ref{contour}, we overlay the contours for SN Ia production in the
($\log P^{\rm i}, M_{\rm 2}^{\rm i}$) plane for initial WD masses
of 0.60, 0.70, 0.75, 0.80, 0.90, 1.0, 1.1 and 1.2 $M_{\odot}$.
Note that the enclosed region almost vanishes when $M_{\rm
WD}^{\rm i}=0.565 M_{\odot}$, which therefore sets the minimum WD
mass for which SD model can produce a SN Ia. We wrote the contours leading to
SNe Ia into a FORTRAN code, which can be used for population synthesis studies.
One can contact X. Meng to request the code.

\section{BINARY POPULATION SYNTHESIS}\label{sect:4}
Adopting the results in Section \ref{sect:3},  we have studied the
supernova frequency from the SD channel via the rapid binary
evolution code developed by \citet{HUR00, HUR02}. Here after, we
use the word {\sl primordial} to represent the binaries before the
formation of SD systems and {\sl initial} for SD systems.

\subsection{Common Envelope}\label{subs:4.1}
Common envelope (CE) is very important for the formation of a SD
systems. We firstly introduce the treatment for CE in this
paper. During binary evolution, the primordial mass ratio
(primary to secondary) is crucial for the first mass transfer. If
it is larger than a critical mass ratio, $q_{\rm c}$, the first
mass transfer is dynamically unstable and a CE forms
(\citealt{PAC76}). The ratio $q_{\rm c}$ varies with the
evolutionary state of the primordial primary at the onset of RLOF
(\citealt{HW87}; \citealt{WEBBINK88}; \citealt{HAN02};
\citealt{POD02}; \citealt{CHE08}). In this study, we adopt $q_{\rm
c}$ = 4.0 when the primary is on MS or crossing HG. This value is supported
by detailed binary evolution studies (\citealt{HAN00};
\citealt{CHE02, CHE03}). If the primordial primary is on FGB or
AGB, we use

\begin{equation}
q_{\rm c}=[1.67-x+2(\frac{M_{\rm c1}^{\rm P}}{M_{\rm 1}^{\rm
P}})^{\rm 5}]/2.13,  \label{eq:qc}
  \end{equation}
where $M_{\rm c1}^{\rm P}$ is the core mass of primordial primary,
and $x={\rm d}\ln R_{\rm 1}^{\rm P}/{\rm d}\ln M_{\rm 1}^{\rm p}$
is the mass-radius exponent of primordial primary and varies with
composition. If the mass donors (primaries) are naked helium
giants, $q_{\rm c}$ = 0.748 based on equation (\ref{eq:qc}) (see
\citealt{HUR02} for details).

Embedded in the CE are the dense core of the primordial primary
and the primordial secondary. Due to frictional drag with the
envelope, the orbit of the embedded binary decays and a large part
of the orbital energy released in the spiral-in process is
injected into the envelope (\citealt{LS88}). Here, we assume that
the CE is ejected if
\begin{equation}
\alpha_{\rm CE}\Delta E_{\rm orb}\geq |E_{\rm bind}|,
  \end{equation}
where $\Delta E_{\rm orb}$ is the orbital energy released, $E_{\rm
bind}$ is the binding energy of common envelope, and $\alpha_{\rm
CE}$ is CE ejection efficiency, i.e. the fraction of the released
orbital energy used to eject the CE. Since the thermal energy in
the envelope is not incorporated into the binding energy,
$\alpha_{\rm CE}$ may be greater than 1 (see \citealt{HAN95} for
details about the thermal energy). In this paper, we set
$\alpha_{\rm CE}$ to 1.0 or 3.0.

\subsection{Evolution Channels}\label{subs:4.2}

There are three channels to produce WD + MS systems according to
the situation of the primary in a primordial system at the onset
of the first RLOF.

\emph{Case 1} (\emph{He star channel}): the primordial primary is in HG or on
RGB at the onset of the first RLOF (i.e. case B evolution defined
by \citealt{KW67}). In this case, a CE is formed because of a
large mass ratio or a convective envelope of the mass donor. After
the CE ejection (if it occurs), the mass donor becomes a helium
star and continues to evolve. The helium star likely fills its
Roche lobe again after the central helium is exhausted. Since the
mass donor is much less massive than before, this RLOF is
dynamically stable, resulting in a close CO WD+MS system (see
\citealt{NOM99, NOM03} for details).

\emph{Case 2} (\emph{EAGB channel}): the primordial primary is in early
asymptotic giant branch stage (EAGB) (i.e. helium is exhausted in
the core, while thermal pulses have not yet started). A CE is
formed because of dynamically unstable mass transfer. After the CE
is ejected, the orbit decays and the primordial primary becomes a
helium red giant (HeRG). The HeRG may fill its Roche lobe and
start the second RLOF. Similar to the He star channel, this RLOF
is stable and produces WD + MS systems after RLOF.

\emph{Case 3} (\emph{TPAGB channel}): the primordial primary fills its Roche
lobe at the thermal pulsing AGB (TPAGB) stage. Similar to the
above two channels, a CE is formed during the RLOF. A CO WD + MS
binary is produced after CE ejection.

However, only one channel can form WD + RG systems and then produce SNe Ia, i.e.
the TPAGB channel. After a WD + MS system is produced from the TPAGB,
the companion continues to evolve
until RG stage. Then, a WD + RG system forms.

The SD systems continue to evolve and the secondaries may
also fill their Roche lobes at a stage and Roche lobe overflow
(RLOF) starts. We assume that if the initial orbital period,
$P_{\rm orb}^{\rm i}$, and the initial secondary mass, $M_{\rm
2}^{\rm i}$, of a SD system locate in the appropriate regions
in the ($\log P^{\rm i}, M_{\rm 2}^{\rm i}$) plane for SNe Ia at
the onset of RLOF, a SN Ia is then produced.

\subsection{Basic Parameters in Monte Carlo Simulation}\label{subs:4.4}
To investigate the birth rate of SNe Ia, we followed the evolution
of $10^{\rm 7}$ binaries via Hurley's rapid binary
evolution code (\citealt{HUR00, HUR02}). The results of grid
calculations in section \ref{sect:3} are incorporated into the
code. The primordial binary samples are generated in a Monte Carlo way
and a circular orbit is assumed for all binaries. The basic
parameters for the simulations are as follows.

(i) The initial mass function (IFM) of \citet{MS79} is adopted.
The primordial primary is generated according to the formula of
\citet{EGG89}
\begin{equation}
M_{\rm 1}^{\rm p}=\frac{0.19X}{(1-X)^{\rm 0.75}+0.032(1-X)^{\rm
0.25}},
  \end{equation}
where $X$ is a random number in the range [0,1] and $M_{\rm
1}^{\rm p}$ is the mass of the primordial primary, which ranges
from 0.1 $M_{\rm \odot}$ to 100 $M_{\rm \odot}$.

(ii) The mass ratio of the primordial components, $q$, is a very
important parameter for binary evolution while its distribution is
quite controversial. For simplicity, we take a uniform mass-ratio
distribution (\citealt{MAZ92}; \citealt{GM94}):
\begin{equation}
n(q)=1, \hspace{2.cm} 0<q\leq1,
\end{equation}
where $q=M_{\rm 2}^{\rm p}/M_{\rm 1}^{\rm p}$.

(iii) We assume that all stars are members of binary systems and
that the distribution of separations is constant in $\log a$ for
wide binaries and falls off smoothly at close separation:
\begin{equation}
an(a)=\left\{
 \begin{array}{lc}
 \alpha_{\rm sep}(a/a_{\rm 0})^{\rm m} & a\leq a_{\rm 0};\\
\alpha_{\rm sep}, & a_{\rm 0}<a<a_{\rm 1},\\
\end{array}\right.
\end{equation}
where $\alpha_{\rm sep}\approx0.070$, $a_{\rm 0}=10R_{\odot}$,
$a_{\rm 1}=5.75\times 10^{\rm 6}R_{\odot}=0.13{\rm pc}$ and
$m\approx1.2$. This distribution implies that the numbers of wide
binary system per logarithmic interval are equal, and that
approximately 50\% of the stellar systems are binary systems with
orbital periods less than 100 yr (\citealt{HAN95}).

(iv)We simply assume a single starburst (i.e. $10^{\rm 11}
M_{\odot}$ in stars are produced one time) or a constant star
formation rate $S$ (SFR) over the last 15 Gyr. The constant SFR is
calibrated so that one binary with $M_{\rm 1}>0.8 M_{\odot}$ is
born in the Galaxy each year (see \citealt{IT84}; \citealt{HAN95};
\citealt{HUR02}). From this calibration, we can get $S=5$ $M
_{\odot}$ ${\rm yr^{-1}}$ (see also \citealt{WK04}). The constant
star formation rate is consistent with the estimation of
\citet{TIM97}, which successfully reproduces the $^{26}$Al
1.809-MeV gamma-ray line and the core-collapse supernova rate in
the Galaxy (\citealt{TIM97}). Actually, a galaxy would have a
complicated star formation history, while the two choices here are
extremes for simplicity. A constant SFR is a good approximation
for spiral galaxies like Galaxy (\citealt{ROCHA00a, ROCHA00b}),
while a single starburst is for elliptical galaxies.

\section{The RESULTS of BINARY POPULATION SYNTHESIS}\label{sect:5}
\subsection{the Birth rates of SNe Ia}\label{subs:5.1}
 \begin{figure}
 \centerline{\includegraphics[angle=270,scale=.35]{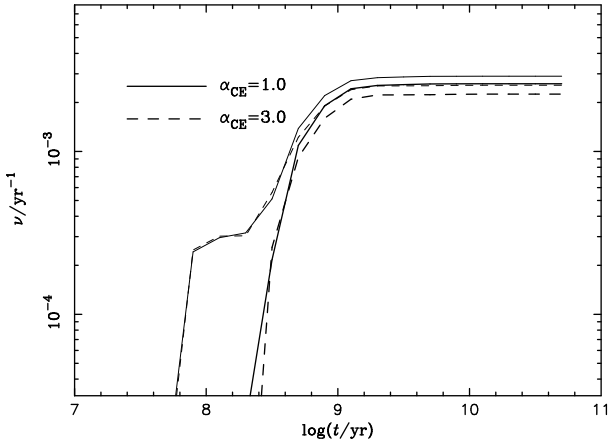}}
 \caption{The evolution of the birth rates of SNe Ia for a constant
 star formation rate (Z=0.02, SFR=$5 M_{\rm \odot}{\rm yr^{\rm -1}}$). Solid and dashed lines show the cases with $\alpha_{\rm
 CE}=1.0$ and $\alpha_{\rm CE}=3.0$, respectively. The thick lines are
 the results in this paper, while the thin lines are the results including the WD + He star channel
 from \citet{WANGB09b}.}\label{sfr}
 \end{figure}

Fig. \ref{sfr} shows Galactic birth rates of SNe Ia for the SD
channel. The simulations give a Galactic birth rate of
$2.25-2.6\times10^{\rm -3}$ yr$^{\rm -1}$ (thick lines), only
slightly lower than the birth rate inferred observationally
(3-4$\times10^{\rm -3}{\rm yr^{\rm -1}}$, \citealt{VAN91};
\citealt{CT97}). If WD + He star channel is included
(\citealt{WANGB09b}), the Galactic birth rates of SNe Ia for the
SD channel becomes $2.55-2.9\times10^{\rm -3}$ yr$^{\rm -1}$ (thin
lines). The WD + He star channel contributes to SNe Ia by about 10 per cent.

\begin{figure}
\centerline{\includegraphics[angle=270,scale=.35]{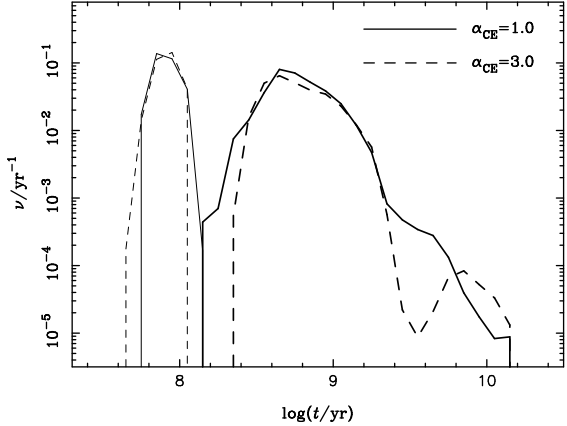}}
\caption{The evolution of the birth rates of SNe Ia for a single
starburst of $10^{\rm 11}M_{\odot}$ for different $\alpha_{\rm CE}$
(solid lines: $\alpha_{\rm CE}=1.0$; dashed lines: $\alpha_{\rm CE}=3.0$).
The thick lines are the results in this paper, while the thin lines are the results from the WD + He star channel
from \citet{WANGB09b}.}\label{single}
\end{figure}

Fig. \ref{single} displays the evolution of birth rates of SNe Ia
for a single starburst of $10^{\rm 11}$ $M_{\odot}$.
Most of the supernova explosions occur between
0.15 and 2.5 Gyr after the burst, which may mean that S/S0 galaxies with ages
below 3 Gyr have a higher SNe Ia birth rate than those older galaxies
(see observational results from \citealt{GALLAGHER08}). Similar to \citet{HAN04}, we also found
that a high $\alpha_{\rm CE}$ leads to a systematically later explosion time,
because a high $\alpha_{\rm CE}$ leads to wider WD binaries,
and, as a consequence, it takes a longer time for the secondary to
evolve to fill its Roche lobe. Actually,
because of the low binding energy of the common
envelope and a long primordial orbital period, $\alpha_{\rm CE}$
has a remarkable influence on CO + WD systems from the TPAGB
channel. Generally, if a CE can be ejected, a low $\alpha_{\rm CE}$
produces a shorter orbital-period WD + MS system,
which is more likely to fulfill the conditions for SNe Ia.
Therefore, we see obvious contribution
from the TPAGB channel when $\alpha_{\rm CE}=1.0$,
but no contribution from this channel
when $\alpha_{\rm CE}=3.0$ (see \citealt{MENG09} for
details regarding the influence of $\alpha_{\rm CE}$ on the TPAGB channel).

In Fig. \ref{single}, it is clear that there is
a high delay-time tail extending to about 15 Gyr, since
the WD + RG system is also included
in this paper. However, the contribution of
WD + RG channel to SNe Ia is much smaller
than that of WD + MS channel. The WD + RG
channel mainly contributes to SNe Ia with delay time
longer than 6 Gyr.

Although \citet{HKN08} claimed that the model with mass-stripping
effect may produce very young population of SNe Ia, there is no
SNe Ia with delay time shorter than $10^{\rm 8}$ yr in our BPS results.
This result is determined by binary evolution, i.e. primordial
primary must be more massive than primordial secondary. Let us
consider a simple example. For a CO WD of $M_{\rm WD}^{\rm i}=1.1$
$M_{\odot}$, its progenitor mass is $\sim6$ $M_{\odot}$
(\citealt{UME99}; \citealt{MENG08}). Then, the upper limit of
primordial secondary mass for an initial CO WD binary system with
$M_{\rm WD}^{\rm i}=1.1$ $M_{\odot}$ is $\sim6$ $M_{\odot}$. When
primordial primary evolves to TPAGB phase, a CE forms because of
dynamically unstable mass transfer. Embedded in the CE are the
dense core of the primordial primary and the primordial secondary.
However, if the primordial secondary is only slightly less massive
than the primordial primary, the CE may obtain enough energy to eject
itself even though the orbit of the new binary decays slightly. Then,
it is difficult for the system to fulfill the conditions for SNe
Ia. On the contrary, because the orbital period of a system with
a lower massive secondary must shrink heavily to eject CE, the system may fulfill
the condition for SNe Ia more likely.
Actually, almost all of the primordial secondaries leading to SN Ia have
a mass less than 4.5 $M_{\odot}$ (see Section \ref{subs:5.2.2}).

\begin{figure}
\centerline{\includegraphics[angle=270,scale=.35]{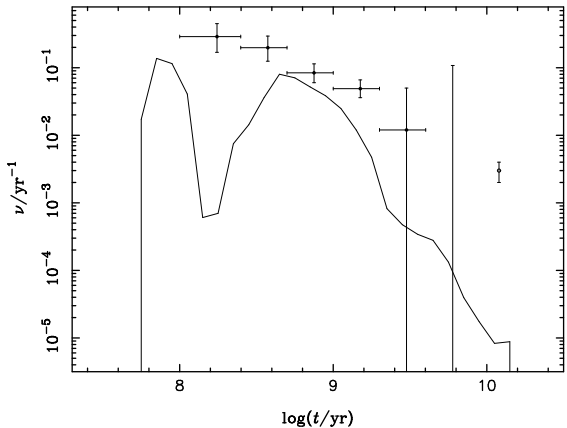}}
\caption{The evolution of the birth rates of SNe Ia for a single
starburst of $10^{\rm 11}M_{\odot}$ for $\alpha_{\rm CE}=1.0$ including WD + He star (\citealt{WANGB09b}), WD + MS
and WD + RG channel. The points are from \citet{TOTANI08} and the crosses represent error bar.}\label{doucom}
\end{figure}

As noted by \citet{WANGB09b}, WD + He star channel may contribute to
very young population of SNe Ia, possible all young population.
Using the results in this paper
and in \citet{WANGB09b}, we constructed the evolution of
the birth rates of SNe Ia for a single
starburst of $10^{\rm 11}M_{\odot}$ for $\alpha_{\rm CE}=1.0$,
shown in Fig. \ref{doucom}.
The constructed distribution of delay time (DDT)
is much similar to that derived from observations by \citet{MAN06},
except that the peak value of young population
is smaller than that in \citet{MAN06}. The WD + He star channel
produces 10 per cent of all SNe Ia, which is the
weak bimodality as suggested by \citet{MANNUCCI08}.
In the figure, we also showed the observational results from \citet{TOTANI08}.
The results in this paper seem to be slightly smaller than those in
\citet{TOTANI08}.
However, considering the large error of observation in \citet{TOTANI08},
our results are not inconsistent with observations at least.

\subsection{Distribution of Initial Parameters of SD Systems for SNe Ia}\label{subs:5.2}
 Observationally, some SD systems are possible progenitors of
 SNe Ia (see the review of \citealt{PAR07}). Further studies are
 necessary to finally confirm them (from both observations and
 theories). In this section, we will present some properties of
 initial SD systems for SNe Ia , which may help us to search
 for the potential progenitors of SNe Ia in the Galaxy.

\subsubsection{Distribution of Initial Masses of WDs}\label{subs:5.2.1}
\begin{figure}
\centerline{\includegraphics[angle=270,scale=.35]{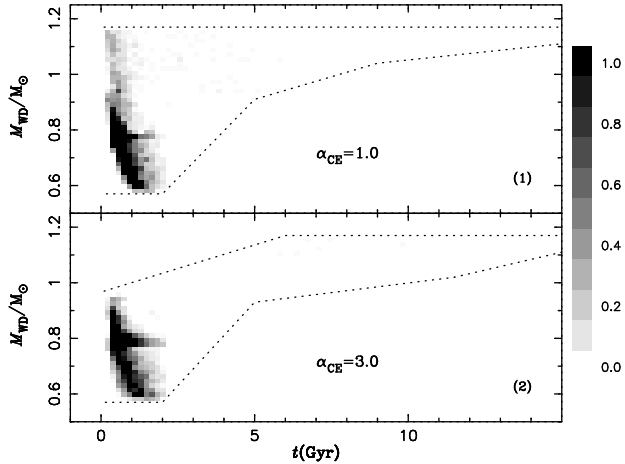}}
\caption{The evolution of the distributions of the initial WD masses in SD
systems for different $\alpha_{\rm CE}$, where a single starburst is assumed.
The dotted lines shows the evolution of the upper and lower boundary of the distributions.
}\label{mwdage}
\end{figure}

In Fig. \ref{mwdage}, we show the evolution of the distribution of the initial WD masses in SD
systems for different $\alpha_{\rm CE}$. We also show the evolution of the upper and lower boundary of the distribution.
From the figure, we can see that most of WDs explodes as SNe Ia in 2 Gyr later after starburst,
where WD + MS systems are the main contributor. $\alpha_{\rm CE}$
does not significantly affect the evolution of the distribution except for WD + MS systems with
age less than 2 Gyr. As shown in \citet{MENG09}, a high $\alpha_{\rm CE}$, i.e. $\alpha_{\rm CE}=3.0$, leads to the disappearing
of the WD + MS system from TPAGB channel. So, we can not see high-mass WDs for $\alpha_{\rm CE}=3.0$
 when age is less than 2 Gyr. For the case of $\alpha_{\rm
 CE}=3.0$, almost all the high-mass WDs are from WD + RG channel.
 There is a clear trend that the range of the WD mass decreases with age,
 while the mean value of WD mass increases. We will discuss this in the Section \ref{sect:6}.

 \subsubsection{Distribution of Initial Secondary Masses}\label{subs:5.2.2}
\begin{figure}
\centerline{\includegraphics[angle=270,scale=.35]{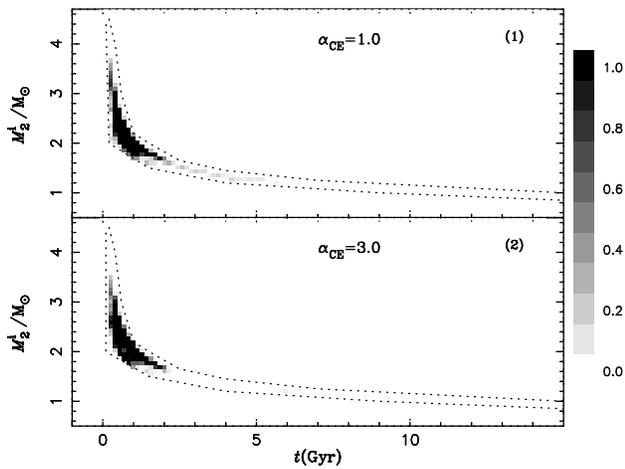}}
\caption{The evolution of the distributions of the initial secondary masses in SD
systems for different $\alpha_{\rm CE}$, where a single starburst is assumed.
The dotted lines shows the evolution of the upper and lower boundary of the distributions.
}\label{mmsage}
\end{figure}

Fig. \ref{mmsage} presents the evolution of distributions of the
initial secondary masses for SNe Ia with different $\alpha_{\rm
CE}$. We also show the evolution of the upper and lower
boundary of the distribution in the figure.  It is clear that the
distributions of the initial orbital period and the evolution of
the distribution for different $\alpha_{\rm CE}$ are similar. The
upper limit of the secondary mass is $4.5 M_{\odot}$, lower than $5 M_{\odot}$
obtained from binary evolution calculation, which indicates that the mass-stripping
effect can not contribute to the young population of SNe Ia.

\subsubsection{Distribution of initial orbital periods}\label{subs:5.2.3}
\begin{figure}
\centerline{\includegraphics[angle=270,scale=.35]{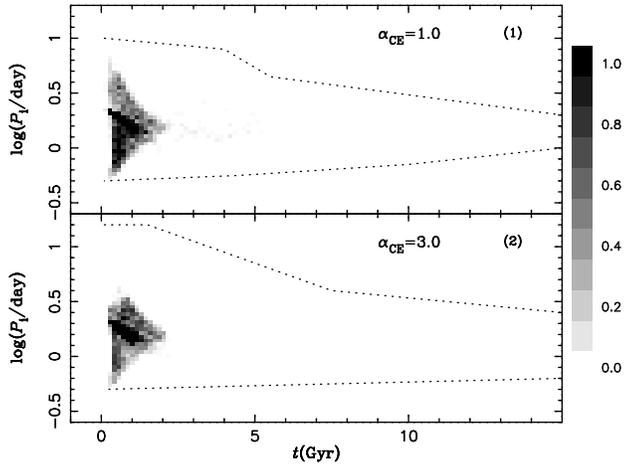}}
\caption{The evolution of the distributions of the initial orbital periods in SD
systems for different $\alpha_{\rm CE}$, where a single starburst is assumed.
The dotted lines shows the evolution of the upper and lower boundary of the distributions.
}\label{perage}
\end{figure}

Fig. \ref{perage} presents the evolution of distributions of the initial
orbital period for SNe Ia for different $\alpha_{\rm CE}$.
We also show the evolution of the upper and lower boundary of the distributions
in the figure. It is clear that the distributions of the initial orbital period and
the evolution of the distributions for different $\alpha_{\rm CE}$ are similar.

 \section{DISCUSSIONS}\label{sect:6}

 \begin{figure}
 \centerline{\includegraphics[angle=270,scale=.35]{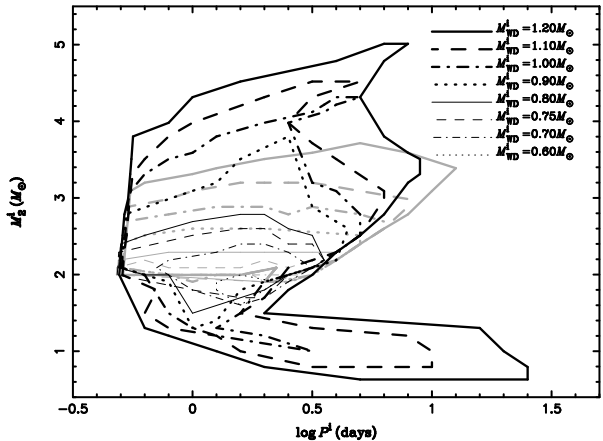}}
 \caption{A comparison of the results in this paper with those
 obtained in \citet{MENG09}. Dark contours show the parameter regions in
 the ($\log P^{\rm i}, M_{\rm 2}^{\rm i}$) plane for different WD masses that lead to a SN Ia
 in this paper. The light grey contours are taken from \citet{MENG09}.
}\label{compare}
 \end{figure}

\subsection{Comparisons with Previous Studies}\label{subs:6.1}
Previous studies show that the minimum mass of CO WDs leading to
SNe Ia, $M_{\rm WD}^{\rm min}$, may be as low as $0.70 M_{\odot}$
for $Z=0.02$ (\citealt{LAN00}; \citealt{HAN04}), and this minimum
mass even heavily depends on metallicity (\citealt{MENG09}). In
this paper, the minimum mass is as low as $0.565 M_{\odot}$, which is much lower
than $0.70 M_{\odot}$. This is because that the mass-stripping effect and the effect of thermal instable disk
are incorporated in our study.
Recently, \citet{WANGB09c} carried out a series of very similar
binary evolution calculation to this paper and obtained
some much similar results, especially for the
WD + RG channel which are almost the same as those in this paper. In their paper, the minimum mass
limit is about 0.61 $M_{\odot}$, higher than that in this paper. This difference
results from the mass-stripping effect considered in this paper.

In Fig. \ref{compare}, we show a comparison of the results in this
paper with those obtained in \citet{MENG09}. Our results are much
different from those in \citet{MENG09}, even only for WD + MS
channel. In \citet{MENG09}, their upper limit for donor stars of
$3.7 M_{\odot}$ in the progenitor binaries is substantially
smaller than the limit of 5.0 $M_{\odot}$ obtained here, which arises from the mass-stripping effect.
The lower limit for donor stars in \citet{MENG09} is also much higher than those in this paper,
which results from the thermal instable disk model.
The upper limits of donor stars for various
initial WD mass in this paper are even higher than those in
\citet{WANGB09c}, which arises from the consideration of
the mass-stripping effect in this paper.

In addition, WD + RG channel are also considered in this
paper. In our results, there is no WD + RG system with period as long as
$10^{\rm 2}$ - $10^{\rm 3}$ days as indicated in \citet{LI97} and \citet{NOM99, NOM03}
(\citet{WANGB09c} obtained a similar result to ours). This is because if the period of an initial
binary system is too long, the mass transfer rate between WD and donor star is too high and
optically thick wind will occur and take much hydrogen-rich material away from the binary system.
The donor star then has no enough material to accumulate on to WD.

 \begin{figure}
 \centerline{\includegraphics[angle=270,scale=.35]{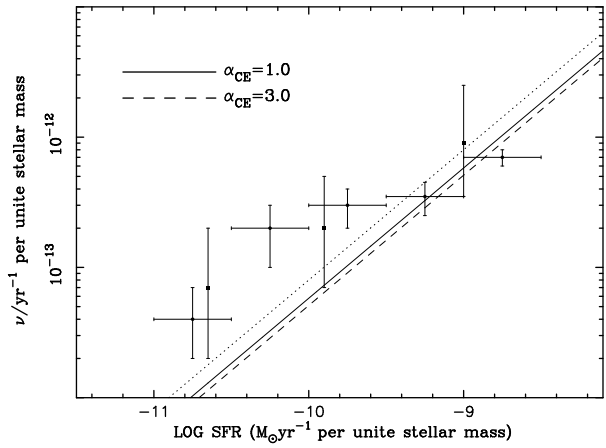}}
 \caption{SNe Ia rate per unit stellar mass as a function of the SFR per
 unit stellar mass for various $\alpha_{\rm CE}$. The WD + He star
 channel in \citet{WANGB09b} is included. The dotted line represents an example
 normalized by Galactic birth rates of SNe Ia. The data are
 obtained from \citet{SULLIVAN06} and the cross represent the error bar.
 }\label{sfrobv}
 \end{figure}

\subsection{Other Possible Channels and Mechanisms for SNe Ia}\label{subs:6.2}
Fig. \ref{sfr} shows Galactic birth rates of SNe Ia for the SD
channel. The simulations give a Galactic birth rate of $2.55 -
2.9\times10^{\rm -3}$ ${\rm yr^{\rm -1}}$ (including WD + He star channel),
which is almost consistent with but seems to be slightly lower
than the birth rate inferred observationally.
Our results are also slightly smaller than those in \citet{TOTANI08}.
In Fig. \ref{sfrobv}, we compared our results (including WD + He star channel)
with the observation from \citet{SULLIVAN06}. We obtained a similar trend to that in
\citet{SULLIVAN06}, i.e. the birth rate of SNe Ia is proportional to SFR. However,
the comparison also shows a slightly lower result than observation, while the example
normalized by Galactic birth rates of SNe Ia may be well consistent with the observation in
\citet{SULLIVAN06}. Therefore, there may be other channels or mechanisms contributing to SNe Ia.

Although, the WD + RG system is included here,
it is still should be studied carefully in the future,
since this channel may explain some SNe Ia with long dealy time.
In addition, RS Oph and T CrB (recurrent
nova, belonging to WD + RG) are both suggested
to be candidates of SNe Ia progenitor
(\citealt{HAC99b}; \citealt{HK06b, HKL07}),
while have a much longer orbital period than
those in this paper (\citealt{LINES88};
\citealt{DOBRZYCKA94, DOBRZYCKA96}).
Perhaps, Symbiotic binaries,
consisting of WD and (super)giant,
not filling its Roche lobe and having stellar wind
should be considered carefully (\citealt{BRANCH95}).
Additionally, the progenitor of some SNe Ia
(e.g. SN 2006X and SN 2007on) are possible WD + RG
systems (\citealt{PAT07}; \citealt{VOSS08}),
although WD + MS channel may also be the candidates
for the progenitor of the SNe Ia (\citealt{HKN08}; \citealt{MYG09a}).

An alternative is the
double-degenerate (DD) channel (\citealt{IT84}; \citealt{WI87}),
although it is theoretically less favored (\citealt{HN00}). In
this channel, two CO WDs with a total mass larger than the
Chandrasekhar mass limit may coalesce and explode as a SN Ia. The
birth rate from this channel is comparable to the observational
value (\citealt{HAN98}; \citealt{YUN98, YUN00}; \citealt{TUT02}),
and SN 2003fg and SN 2005hj likely resulted from the DD channel
(\citealt{HOW06}; \citealt{BRA06}; \citealt{QUI07}).
Observationally, a large amount of DD systems are discovered
(\citealt{NAPIWOTZKI04}), but only KPD 1930+2752 is a possible
progenitor candidate for a SN Ia via DD channel (\citealt{GEI07}).
The total mass of KPD 1930+2752 ($\sim1.52 M_{\rm \odot}$) exceeds
the Chandrasekhar mass limit and the time scale of coalescence is
about 200 Myr estimated from orbital shrinkage caused by
gravitational wave radiation (\citealt{GEI07}). However,
\citet{ERGMA01} argued that, from detailed binary evolution
calculation, the final mass of KPD 1930+2752 is smaller than the
Chandrasekhar mass limit due to a large amount of mass loss during
evolution. In addition, earlier numerical simulations showed that
the most probable fate of the coalescence is an accretion-induced
collapse and, finally,  neutron star formation (see the review by
\citealt{HN00}). A definitive conclusion for DD model is thus
premature at present, and further studies are needed.

In this paper, we have not considered
the influence of rotation on the H-accreting WDs. The
calculations by \citet{YOON04} have shown
that the He-shell burning is much less violent when rotation
is considered. Then, the He-accretion efficiency  may be significantly increase
Meanwhile, the maximum stable mass of a rotating WD may exceed the
Chandrasekhar  mass limit (i.e. the super-Chandrasekharh mass model: \citealt{UENISHI03};
\citealt{YOON05}; \citealt{CHENWC09}).
However, we only focus on the standard Chandrasekhar mass
explosions of the accreting WDs in this paper.

\citet{LIE03, LIE05} and \citet{WF05} found that about 10\% of WDs
have magnetic fields higher than 1MG. The mean mass of these WDs
is 0.93 $M_{\odot}$, compared to mean mass of all WDs which is
0.56 $M_{\odot}$ (see the review by \citealt{PAR07} for details).
{Thus, the magnetic WDs are more likely to reach the Chandrasekhar
mass limit by accretion. Meanwhile, the magnetic field may also
affect some properties of WD+MS systems, e.g. the mass transfer
rate, the critical accretion rate, the thermonuclear reaction rate
etc, leading to a different birth rate of SNe Ia.

\subsection{C/O Ratio: the Origin of the Luminosity Scatter of SNe Ia?}\label{subs:6.3}
It is widely known that there exists a scatter of the maximum
luminosity of SNe Ia and the scatter are affected by their
environment. The most luminous SNe Ia always occur in spiral
galaxies, while both spiral and elliptical galaxies are hosts for
dimmer SNe Ia, which lead to a dimmer mean peak brightness in
elliptical than in spiral galaxies (\citealt{HAM96}). In addition,
the mean peak brightness of SNe Ia in a galaxy has less variation
in the outer regions than in the inner regions (\citealt{WAN97};
\citealt{RIE99}). To explain these phenomena, \citet{NOM99, NOM03}
suggested that the ratio of carbon to oxygen (C/O) of an initial
CO WD is the origin of the luminosity scatter of SNe Ia. The ratio
is a function of WD mass, i.e. a massive CO WD leads to a lower
C/O ratio, and thus a lower amount of $^{\rm 56}{\rm Ni}$
synthesized in the thermonuclear explosion (\citealt{NOM99,
NOM03}), which results in a lower luminosity of SNe Ia
(\citealt{ARN82}; \citealt{ARN85}; \citealt{BRA92}). In Fig.
\ref{mwdage}, we showed the evolution of the range of WD mass with
time and found that the range decreases, while the mean mass of CO
WD increases with time. If the C/O is the origin of the luminosity
scatter of SNe Ia, then our results may well explain the
phenomenon found by \citet{HAM96}. In a spiral galaxy, because of
existing continuous star formation, and then CO WD with various
age and initial mass, the spiral galaxy may be host for luminous
and dim SNe Ia. However, In an elliptical galaxy, no star
formation and no CO WD with young age, only massive CO WD may
contribute to SNe Ia, and then a dimmer SNe Ia is expected. The
different variation of mean peak brightness of SNe Ia between
inner and outer region in a galaxy can also be well explained by
the different WD mass resulting from different metallicity (see
details in \citealt{MENG09}). So, our results and those in
\citet{MENG09} uphold the C/O as the origin of the luminosity
scatter of SNe Ia. Our results in this paper provide a method to
check whether or not the C/O is the origin of the luminosity
scatter of SNe Ia. If the C/O is the origin, the elliptical
galaxies with age less than 2 Gyr should be hosts of SNe Ia with
various luminosity, including very luminous SNe Ia, while
older galaxies are only the host of dimmer SNe Ia. In
other words, age might be the most important factor to determine
the luminosity of SNe Ia, and dimmer SNe Ia would have a wide age
distribution. Interestingly, many observations have obtained similar conclusions
(\citealt{GALLAGHER08}; \citealt{NEILL09}; \citealt{HOWEL09b}).
Considering the mass of secondary decreases with age (see Fig.
\ref{mmsage}), a progenitor with more massive secondary should be
correlated with more luminous explosion (\citealt{HOWEL09b}). In
addition, \citet{HOWEL09b} also noticed that low-stretch (dimmer)
SNe Ia be evenly distributed between host age older than
$5\times10^{\rm 8}$ yr, which is consistent with our results qualitatively.
Our results may also indicate that the mean
maximum luminosity of SNe Ia should increase with redshift.
Interesting, both \citet{HOWELL07} and \citet{SULLIVAN09} found
that the mean `stretch factor' increases with
redshift\footnote{Stretch factor is an indicator of the luminosity
of SNe Ia. A luminous SN Ia usually has a higher stretch factor.}.

\subsection{Young Population of SNe Ia}\label{subs:6.4}
Observationally, there exists very young SNe Ia (younger than
$10^{\rm 8}$ yr). To explain these SNe Ia in the frame of SD model,
\citet{HKN08} designed a WD + MS channel with mass-stripping effect.
However, the detailed BPS study in this paper showed that the
WD + MS channel with mass-stripping effect totally can not produce SNe Ia with very
short delay time. The WD + He star channel can well produce the young SNe Ia (\citealt{WANGB09a, WANGB09b}).
The WD + MS, WD + RG and WD + He star channel can produce a weak bimodality of DDT,
which might mean that there still exist other channel contributing to the young SNe Ia.
Maybe, a symbiotic star with aspherical stellar wind is a possible candidate (\citealt{LGL09}).

 \section{SUMMARY AND CONCLUSIONS}\label{sect:7}
Incorporating mass-stripping effect in \citet{HKN08} and the effect of
the instability of accretion disk on the evolution of WD binaries into Eggleton's stellar evolution code,
and including also the prescription of \citet{HAC99a} for mass-accretion of CO WD,
We performed binary evolution calculations for more than 1600
close WD binaries. Adopting the results obtained here, we carried out a series BPS sutdy
and calculated the birth rate of SNe Ia. We summarize the basic results
as following.

1. The detailed binary evolution calculation results further
confirmed the suggestion in \citet{HKN08} that the mass-stripping
effect may attenuate between WD and the companion, and avoid the
formation of a CE, and then increase the donor mass leading to SNe
Ia. For a reasonable strength of the mass-stripping effect in
physics, i.e. $c_{\rm 1}=1.5$, a companion as massive as 5 $M_{\odot}$
can produce an SNe Ia. However, the detailed BPS study show that
the upper limit of the companion could be $\sim4.5 M_{\odot}$.

2. The detailed binary evolution calculation results confirm that
the disk instability could substantially increase the
mas-accumulation efficiency for accreting WD, and extend the mass
of donor star producing SNe Ia to $\sim 1 M_{\odot}$ (see also
\citealt{XL09} and \citealt{WANGB09c}).

3. CO WDs leading to SNe Ia
may have a mass as low as $0.565 M_{\odot}$.

4. The Galactic birth rate from the SD channel in this paper is
$2.25-2.6\times10^{\rm -3}$ ${\rm yr}^{\rm -1}$, based on
$\alpha_{\rm CE}$. If the WD + He star channel is also included, the
Galactic birth rate from the SD channel increases to
$2.55-2.9\times10^{\rm -3}$ ${\rm yr}^{\rm -1}$, only slightly
smaller than that derived from observation. Then, there should be
other channels or mechanisms contributing to SNe Ia.

5. The DDT from the WD + MS, WD + RG and WD + He star channels may be a weak bimodality.
The WD + MS produces most of SNe Ia with delay time between $0.15-2.5$ Gyr.
The WD + RG channel may mainly contribute to SNe Ia with delay time longer than 6 Gye.

6. The mass-stripping effect may not contribute to very young population of
SNe Ia. The WD + He star channel could contribute most of young SNe Ia, even though not all.

7. Our BPS results may uphold the C/O as the origin of the luminosity scatter of SNe Ia
and provide a method to verify it.

8. We show the evolution of the initial parameters of SD systems for SNe Ia with time,
which is helpful for searching for possible candidates of SNe Ia progenitor
systems.

\section*{Acknowledgments}
This work was supported by Natural
Science Foundation of China under Grant Nos. 10963001.

\end{document}